\documentclass{llncs}

\usepackage{amsbsy}
\usepackage{amssymb}
\usepackage{mathrsfs}
\usepackage{latexsym}
\usepackage{enumerate}
\usepackage{amsmath}
\usepackage{amsfonts}
\usepackage{float}

\usepackage[ruled]{algorithm}

\newcommand{\qedsymb}{\hfill{\rule{2mm}{2mm}}}
\newenvironment{proof1}{\begin{trivlist}
\item[\hspace{\labelsep}{\bf\noindent Proof: }]
}{\qedsymb\end{trivlist}}

\floatname{algorithm}{Protocol}

\begin{document}

\title{\LARGE\bf Information Dissemination in Unknown Radio networks with Large Labels}

\author{Shailesh Vaya}
\institute{E-mail : shailesh.vaya@gmail.com\\Department of Computer Science and Engineering\\ Indian Institute of Technology Patna\\India - 800013}

\maketitle
\abstract{
\noindent We consider the problems of deterministic broadcasting and gossiping in completely unknown ad-hoc radio networks. We assume that nothing is known to the nodes about the topology or even the size of the network, $n$, except that $n > 1$. Protocols for vanilla model, when $n$ is known, may be run for increasingly larger estimates $2^i$ on the size of the network, but one cannot determine when such a protocol should terminate. Thus, to carry this design paradigm, successfull completion or incompletion of the process should be detected, and this knowledge circulated in the network. In radio networks litertaure, this setting is referred to as Acknowledged broadcasting and gossiping. An important feature of dynamic ad-hoc radio networks is that radio nodes become ineffective due to a variety of reasons and new nodes are introduced from time-to-time. Thus, the new nodes have to assigned labels in a much larger range, say polynomial in the size of the network, e.g., $[1,\dots,n^c]$, for some constant $c$.
\newline
  For the above setting, we present the following results for strongly connected networks: (a) A deterministic protocol for acknowledged broadcasting which takes $NRG(n,n^c)$ rounds, where $NRG(n,n^c)$ is the round complexity of deterministic gossiping for vanilla model. (b) A deterministic protocol for acknowledged gossiping, which takes $O(n^2 \lg n)$ rounds when collision detection mechanism is available. The structure of the transmissions of nodes in the network, to enable them to infer collisions, and discover existence of unknown in-neighborhood as a result, is abstracted as a family of integral sets called Selecting-Colliding family. We prove the existence of Selecting-Colliding families using the probabilistic method and employ them to design protocol for acknowledged gossiping when no collision detection mechanism is available.

  Finally, we present a deterministic protocol for acknowledged broadcasting for bidirectional networks, with a round complexity of $O(n \lg n)$ rounds.
}

\noindent {\bf Keywords:} Unknown Radio Networks, Acknowledged Broadcasting and Gossiping, Polynomially Large Labels, Selecting-Colliding Families.

\setcounter{page}{1}

\section{Introduction}
\label{sec:introduction}
  Mobile ad-hoc radio networks play an important role in a wide range of fields, ranging from agriculture and automobiles to hazardous environments and defense. Often enough, the connectivity of radio networks is not planned ahead precisely and there is no centralized system to coordinate the deployed radio stations. Thus, radio networks form a classical distributed setting in which the nodes have to rely solely on the communication received by them from other nodes, their own label values (possibly knowledge of some other parameters) and environment stimulus to determine their action at any time. Complex communication activities rely in turn on fundamental communication primitives like broadcasting and gossiping.

  In the {\it broadcasting problem}, a message from a distinguished source node is to be communicated to the rest of the nodes of the network, while in the {\it gossiping} problem each node of the network possesses a message which is to be disseminated to the entire network. Needless to say, gossiping can be achieved only on strongly connected networks. For both these primitives, the most important design consideration is the total time lapsed from initiation of a task to its completion. In this work, we assume that nodes are synchronized with respect to a clock, there are no faulty nodes and that nodes can communicate messages of arbitrary sizes in a single time unit. Computation is organized in rounds. In every round, each node acts either as a transmitter or as a receiver but not both. Whether a node is actually able to receive a message or the message sent by it is received by an out-neighbor depends on the following salient feature of radio networks: {\em If two in-neighbors of a node transmit any message in the same round, then a collision occurs and the receiving node receives nothing. Furthermore, the receiving node cannot distinguish it from the case when all of its in-neighbors are silent}. It is this feature of radio networks that makes even a simple task of broadcasting non-trivial.

  Two metrics by which the complexity of broadcasting and gossiping protocols are usually measured are round complexity and message complexity. These metrics are typically defined in terms of the number of nodes in the network $n$, the diameter of the network, maximum in-degree of a node etc. The most important and prevalent complexity measure under which broadcasting and gossiping problem has been studied is {\it round complexity}, which is the criterion used to calibrate protocols in this work.

  The broadcasting and gossiping problem in radio networks have been studied under different assumptions about the knowledge of topology. In the centralized setting, the topology of the network is known to all the nodes. In a stricter model, nodes are given the labels of their immediate neighbors and the more general setting is one when the network is directed and nodes know only their own labels and the number of nodes in the network. {\it However, the {\it ad-hoc} nature of the radio networks is most closely modeled by the setting in which even the number of nodes in the radio network is not known}. Radio nodes have batteries which run for a short duration and whose life is largely restricted by the number of radio transmissions made by a node. Once the battery is consumed the radio node itself may be discarded or forgotten. Older nodes lying in the field for long duration are likely to die out, may be ignored and stop participating in future computations. In a truly dynamic setting, nodes do not have any estimate of the size of the network to which they belong. Also, many nodes in the field may not respond, may possess small labels which cannot be reused. Thus, the ad-hoc nature of radio networks imposes the constraint that the nodes are allowed to take polynomially large labels i.e., within some range $[1,\dots,n^c]$, for some large constant $c$ (The reader is referred to \cite{P} for a detailed motivation for studying the case of large labels).

  A protocol designed for the vanilla setting can be adapted to work for this model, by running it for multiplicatively larger estimates on the size of the network in every new phase. For this phase-wise design methodology nodes cannot determine when a particular task has completed. Hence, it is required that nodes receive some form of an ACK (acknowledgment) or NACK (negative acknowledgment) about the completion of certain task at the end of a phase. In the literature, these problems for completely {\it unknown networks} are called {\it acknowledged} broadcasting and acknowledged gossiping. They are known to be impossible, to achieve, for arbitrary $n$. If $n$ is guaranteed to be greater than $1$, then possibility results are known, \cite{CGGPR02}, \cite{OCW03}. Secondly, the impossibility result in \cite{FP08} rigorously formalizes the intuition that for acknowledged broadcasting, players must wake-up spontaneously and start participating on their own. Thus, we assume sponatenous transmissions and $n > 1$.

  The question of acknowledged broadcasting and gossiping was initially formulated and explored in \cite{CGGPR02}. Most recently, \cite{UCW07} and \cite{FP08} considered this problem. \cite{UCW07} give protocols for acknowledged broadcasting (ARB) and acknowledged gossiping (ARG) for bidirectional and strongly connected networks. For $ARB$ on bidirectional and strongly connected networks, \cite{UCW07} present protocols of round complexity $O(n)$ and $\tilde{O}(n^{4/3})$, respectively. For ARG, the authors design protocols of round complexity $O(n \lg^3 n)$ and $\tilde{O}(n^{4/3})$ for bidirectional and strongly connected networks, respectively. For strongly connected networks, \cite{FP08} improve the round complexity for deterministic acknowledged broadcasting to $O(n \lg n \lg \lg n)$ rounds.

  Acknowledged protocols proposed in \cite{CGGPR02}, \cite{OCW03}, \cite{UCW07} and \cite{FP08}, for the setting of small labels, are very inefficient when the nodes can take polynomially large labels; and not much useful in practice. From a theoretical point of view, the assumption of small labels, allows protocol designers to use a simple round-robin phase where all nodes transmit their IDs, in separate rounds corresponding to their labels and learn about their in-neighborhoods etc.. This considerably simplifies the construction of these protocols. However, when nodes can take large labels, a round-robin phase requires $\Omega(n^c)$ rounds as node only transmits when the round corresponding to its label has reached. In this work, we design efficient deterministic protocols for radio networks for acknowledged broadcasting and gossiping, improving upon all previous results when nodes can be assigned (polynomiall) large labels.

  An overview of the results in this work is presented in Section \ref{sec:discussion}.

\subsection{Related Works}
\label{subsec:related-works}
  The study of broadcasting has received considerably more attention then gossiping. For directed graphs, a series of works improved the trivial upper bound of $O(n^2)$ for broadcasting (when nodes have small labels), to $O(n^{11/16})$ in \cite{CGGPR02}, $\tilde{O}(n^{5/3})$ in \cite{MP01}, $O(n^{3/2})$ in \cite{CGOR00}, $O(n \lg^2 n)$ in \cite{CGR02}, $O(n \lg_D^2)$ in \cite{CR} and most recently $O(n \lg n \lg \lg n)$ in \cite{M10}. \cite{BP} show that for any deterministic algorithm $\mathcal{A}$ for broadcasting in ad-hoc radio networks, there are networks on which $\mathcal{A}$ requires $\Omega(n \lg n)$ rounds.

  The first sub-quadratic deterministic algorithm for the gossiping problem in ad-hoc radio networks was $\tilde{O}(n^{3/2})$ time algorithm proposed in \cite{CGR02}. Subsequently, \cite{Xu} improved this bound by a poly-logarithmic factor obtaining $O(n^{3/2})$ bound. For small diameter $D$, the gossiping time was later improved in \cite{GL02} to $\tilde{O}(nD^{1/2})$. These algorithms assume that the node labels are linear in $n$. \cite{CMS} present a $\tilde{O}(D {\Delta}^2)$-time deterministic algorithm, which was improved to $\tilde{O}(D \Delta^{3/2})$ in \cite{GL02}, for polynomially large labels. \cite{GRX04} improved the result on deterministic gossiping from $\tilde{O}(n^{5/3})$ in \cite{GPP02} to $\tilde{O}(n^{4/3})$ in \cite{GRX04}. An excellent survey of results on deterministic gossiping problem is presented by Gasieniec in \cite{G09}. For bounded size messages deterministic gossiping was studied in \cite{CGL02} and for unit size messages gossiping protocols were given in \cite{GP} and \cite{MX}. \cite{CGR02} propose an $O(n \lg^4 n)$ time randomized gossiping algorithm, which was improved to $O(n \lg^3 n)$ rounds in \cite{LP} and $O(n \lg^2 n)$ rounds in \cite{CR}.

  \cite{CGGPR02}, \cite{OCW03}, \cite{UCW07} study the problem of acknowledged broadcasting and gossiping. \cite{FP08} improve a result in \cite{UCW07} and present a $O(n \lg n \lg \lg n)$-time deterministic algorithm for acknowledged broadcasting. As pointed above the shortcoming of these works is that the nodes can take small labels only.

  For the centralized setting, the topology of the radio network is known to all the nodes of the network. For this setting, a deterministic broadcasting protocol of $O(D \lg^2 n)$ is given in \cite{CW}, for networks of radius $D$. For the centralized setting, a lower bound of $\Omega(D + \lg^2 n)$ rounds was proved in \cite{ANLP}. This was shown to be tight in a recent work \cite{KP5}, where a matching upper bound of $O(D + \lg^2 n)$ rounds was proved, after a series of improvements from $O(D \lg^2 n)$ rounds in \cite{CW}), $O(D \lg n + \lg^2 n)$ in \cite{KP4}, $O(D + \lg^4)$ in \cite{EK} and $O(D + \lg^5 n)$ in \cite{GM}.

\subsection{Organization of the paper}
\label{subsec:organization}
  Section \ref{sec:model-and-tools} gives a formal description of the model, some relevant tools and terminologies. An overview of the results in this work is given in section \ref{sec:discussion}. A deterministic protocol for acknowledged broadcasting in strongly connected networks is presented in section \ref{sec:arbllstsl}. Deterministic protocol for acknowledged gossiping in strongly connected networks when collision detection is available to the nodes of the network is presented in section \ref{sec:gossip}. A new combinatorial structure called {\it Selecting Colliding family} is defined and shown to exist in section \ref{selectingcolliding}. Protocol for acknowledged gossiping on strongly connected networks, using Selecting-Colliding families, without collision detection mechanism, is given in section \ref{sec:NoCdProt}. Section \ref{sec:arbstbl}, presents a deterministic protocol for acknowledged broadcasting for bidirectional networks.

\section{Preliminaries}
\label{sec:model-and-tools}
  A protocol $\pi$ for a radio network $N$ is a synchronous multi-processor protocol with the following salient properties: (1) Time is slotted in rounds. (2) The topology and size of the network $n$ is unknown. Each node is assigned a unique label $l \in [1,\dots,n^c]$, where $c$ is some pre-defined constant. (3) All nodes execute identical copies of the same protocol $\pi$. (3) In each round, every node either acts as a transmitter or as a receiver (or is inactive). (4) A node receives a message in a specific round if and only if it acts as a receiver and exactly one of its neighbors transmits in that round. If particular, if two or more in-neighbors of a node transmit in the same round, then the node receives $\phi$, which is what it receives when none of its in-neighbors transmit. We assume that the messages are authenticated, that is, when a node receives a message it gets to know the label of the transmitting node.  (5) The action of a node in a specific round is determined by \begin{enumerate} \item Initial input, which typically consists of its own label and constant $p$, which is presumed to be known.  \item Messages received by the node in previous rounds.  \item Current round number, which is a global variable known to all nodes of the network. \end{enumerate} (7) $n \geq 2$ (It is known from the result in \cite{CGGPR02} that acknowledged broadcasting is impossible when $n=1$). (8) Spontaneous transmissions are allowed i.e., nodes can start transmitting without having received any message (It is known from the result in \cite{FP08}, that acknowledged broadcasting is impossible when spontaneous transmissions are not allowed).

\subsection{Combinatorial tools}
\label{sec:tools}
  We review a few standard combinatorial tools in radio networks literature which are used in this work.

\begin{definition} (Selective family)
\label{selectivefamily}
  A ($k,m$)-selective family $\mathcal{F}$ consists of a set of subsets of a set $S = [1,\dots,m]$, such that for every subset $S_k$ of $S$ of size at most $k$, there exists at least one subset $S_f \in \mathcal{F}$ for which $|S_f \bigcap S_k| = 1$. The size of a ($k,m$)-selective family $\mathcal{F}$ is referred by $SF(k,m)$.
\end{definition}
\begin{theorem}
\label{theo:selectivefamily}
  There exists a ($k,m$)-selective family $\mathcal{F}$ of size $O(k \cdot \lg \frac{m}{k})$, \cite{CMS}.
\end{theorem}

  By execution of a selective family it is meant that

\begin{definition} (Strongly Selective family)
\label{stronglyselectivefamily}
  $\mathcal{F}$ is a ($k,m$)-strongly selective family if for every subset $S_k$ of $S$ of size at most $k$, and every member $z_k \in S_k$, there exists at least one subset $S_f \in \mathcal{F}$ for which $S_f \bigcap S_k = \{ z_k \}$, \cite{CMS}. The size of a ($k,m$)-Strongly Selective family $\mathcal{F}$ is referred by $SSF(k,m)$.
\end{definition}

\begin{theorem}
\label{theo:stronglyselectivefamily}
  There exists a ($k,m$)-strongly selective family $\mathcal{F}$ of size $O(k^2 \cdot \lg \frac{m}{k})$, \cite{EFF}.
\end{theorem}

  The fastest known protocol for broadcasting in directed networks, is from \cite{M10}:
\begin{theorem}
\label{fastbroadcast}
  There exists a deterministic broadcasting protocol $RB(n,n^c)$ that works in $O(n \lg n \lg \lg n)$ rounds on all directed graphs. When $n$ nodes of the network can take labels in $[1, \dots, n^c]$.
\end{theorem}

  The number of rounds taken by $RB(n,n^c)$ is referred by $NB(n,n^c)$.\\

  The fastest known protocol for deterministic gossiping in strongly connected networks is from \cite{GRX04}:
\begin{theorem}
\label{fastgossip}
  There exists a deterministic gossiping protocol, $RG(n,n^c)$, that works in $\tilde{O}(n^{4/3})$ rounds on all strongly connected networks with $n$ nodes, when nodes can have labels in range $[1,\dots,n^c]$.
\end{theorem}

  The number of rounds taken by $RG(n,n^c)$ is referred by $NRG(n,n^c)$.

\subsection{Notations}
\label{notations}
  We denote the ider the problems of deterministic broadcasting and gossiping in completely unknown ad-hoc radio networks. We assume that nothing is known to the nodes about the topology or even the size of the network, $n$, except that $n > 1$. Protocols for vanilla model, when $n$ is known, may be run for increasingly larger estimates $2^i$ on the size of the network, but one cannot determine when such a protocol should terminate. Thus, to carry this design paradigm, successfull completion or incompletion of the process should be detected, and this knowledge circulated in the network. In radio networks litertaure, this setting is referred to as Acknowledged broadcasting and gossiping. An important feature of dynamic ad-hoc radio networks is that radio nodes become ineffective due to a variety of reasons and new nodes are introduced from time-to-time. Thus, the new nodes have to assigned labels in a much larger range, say polynomial in the size of the network, e.g., $[1,\dots,n^c]$, for some constant $c$.
\newline
  For the above setting, we present the following results for strongly connected networks: (a) A deterministic protocol for acknowledged broadcasting which takes $NRG(n,n^c)$ rounds, where $NRG(n,n^c)$ is the round complexity of deterministic gossiping for vanilla model. (b) A deterministic protocol for acknowledged gossiping, which takes $O(n^2 \lg n)$ rounds when collision detection mechanism is available. The structure of the transmissions of nodes in the network, to enable them to infer collisions, and discover existence of unknown in-neighborhood as a result, is abstracted as a family of integral sets called Selecting-Colliding family. We prove the existence of Selecting-Colliding families using the probabilistic method and employ them to design protocol for acknowledged gossiping when no collision detection mechanism is available.

  Finally, we present a deterministic protocol for acknowledged broadcasting for bidirectional networks, with a round complexity of $O(n \lg n)$ rounds.
ower set of set $S$ by $2^{S}$.\\

  We design protocols for $Task(n,n^c)$, where the Task is either broadcasting or gossiping, $n$ denotes the number of nodes in the network and $n^c$ denotes an upper bound on the value of the largest label of a node in the network.\\

  Our protocols for acknowledged broadcasting and gossiping proceed in phases $i = \{4,\dots,\lg n^c\}$. In phase $i$, all nodes with labels less than or equal to $2^i$ are called {\it small}, nodes with labels greater than $2^i$ but less than $2^{i \cdot c}$ are called {\it medium} and all nodes with labels greater than $2^{i \cdot c}$ are called {\it big}.

\section{Overview of our results}
\label{sec:discussion}
  We consider the problem of designing deterministic protocols for broadcasting and gossiping on strongly connected networks, when nodes can take large labels. For sake of completeness, the complete set of assumptions about the model of radio networks, considered in this work, are presented in Section \ref{sec:model-and-tools}.

  The general methodology for design of protocols, for (completely) unknown networks i.e., the acknowledged setting, \cite{UCW07}, \cite{FP08} etc., is as follows: Execute the protocol for the vanilla model, in a phase-wise manner, where the $i^{th}$ phase is designed to run a vanilla protocol for estimate $n=2^i$ on the number of (live) nodes in the network. At the end of phase $i$, nodes in the network are supposed to receive a {\it feedback} that the protocol has not completed sucessfully. The alternate, that the protocol has completed successfully, is inferred by hearing {\it silence} in those very rounds, when the {\it failure} feedback is circulated. This feedback is called {\it acknowledgement} and we use this term to discuss our results.

\subsection{Acknowledged broadcasting with large labels}
  For the broadcast protocol, the main difficulty in carrying out the doubling scheme is for the source to determine when the broadcasting/gossiping process has completed. On the execution of broadcast protocol $RB(n,n^c)$ on the network, with current estimate $n$ (on size of network) and $n^c$ (size of maximum label of a node) some nodes receive the message. The source should somehow receive the feedback if there are (a) Nodes in the network which did not receive the message. (2) Nodes whose labels are outside the currently estimated range of $[1, \dots, n^c]$. The latter case is considered, because neighbors of such nodes may not have received the message, as execution of $RB(n,n^c)$ with a low estimate of $n$, does not ensure all nodes in the network to receive the message, in the current run.

  The problem of {\it detecting the failure} in the network is done using a similar technique used in \cite{UCW07}, \cite{FP08}. In \cite{FP08}, who improve the result for broadcasting in \cite{UCW07}, for small labels, this is achieved as follows. Let one or more small labelled nodes, i.e., nodes with label value less than current estimate $n$, become aware of the failure, due to any of the reasons described above. An execution of the broadcast protocol $RB(n+1,d \cdot n)$, (i.e., broadcast protocol which should work for broadcast initiated by potentially multiple nodes, observed by Gasienec and formally proved in \cite{FP08}), amongst the small nodes with labels $< d \cdot n$, the failure message also reaches the source node. This works as only nodes with small labels participate in protocol $RB(n, d \cdot n)$. The source then initiates the execution of the next phase. However, when nodes can have large labels i.e., up to $n^c$, the execution of broadcast protocol $RB(n+1,n^c)$ amongst nodes with labels less than $n^c$, complications arise and this may not be achieved successfully. Specifically, this is because the density of nodes in the reachable network, in range $[1,\dots,n^c]$, may happen to be far larger $O(n)$. This may cause many unpredicted collisions, for e.g., a node with very high indegree of $O(n^c)$ may exist, which may be on the path to the source from the failed node, disallowing successfull conclusion of $RB(n+1,n^c)$. Thus, although failure is detected, the failure message does not reach the source at all. The source, then incorrectly infers the ensuing silence as successfull completion of the broadcast protocol and so on.

  We handle this as follows. The source broadcasts the message using the current estimate on the size of the network $n$ and wakes up different nodes of the network in the process. A gossiping protocol, with current estimates on the size of network, is then executed multiple times, with appropriate messages, amongst the nodes of the network. This enables the source to identify if there is a node, in the network, which is aware of the failure. The source then rebroadcasts this information in the network at the end of the phase, resulting in all nodes proceeding to execution of the next phase of the protocol. We establish the following result.
\begin{theorem}
\label{ack:broadcast}
  There exists a deterministic protocol, which achieves broadcasting on all strongly connected radio networks of $n$ nodes, when nodes can have large labels, in $O(NRG(n,n^c))$ rounds (where $NRG(n,n^c)$ is the round complexity of deterministic radio gossiping for the vanilla model, when nodes can take large labels).
\end{theorem}

\subsection{Acknowledged gossiping with collision detection mechanism}
\label{subsec:ackgpnocd}
  We realize that our protocol for Acknowledged Broadcasting also achieves (or may be adapted to achieve) gossiping amongst all nodes of the strongly connected network. However, the reason it is not a protocol for Acknowledged Gossiping is that it assumes a single source which can act as the leader and direct activities in the network. This is not the case for the Acknowledged Gossiping problem in which all nodes start with the same designation. Leader election is achieved by binary selection on a range of labels and proceeds by repeatedly invoking the broadcast protocol $RB(n+1,n^c)$, on a network of $n$ nodes, with broadcast initiated by multiple nodes (refer to \cite{FP08} for a verbose proof of why this works).

  Realizing this approach explicitly in the current setting of large labels encounters the following problem: $RB(2^i+1,2^{i \cdot c})$ is not guaranteed to complete successfully. Initiators $s_1, s_2, s_3,\dots$ which have labels in a certain range may reach different subsets of nodes $\mathcal{M}^1_i, \mathcal{M}^2_i, \mathcal{M}^3_i, \dots$ etc. in the network, respectively. This is unlike a single set $\mathcal{M}_i$ defined by single-initator broadcast, for which $\overline{\mathcal{M}}_i$ easily categorizes the subset of unreachable nodes. However, for the former case, a member of $\overline{\mathcal{M}^1_i}$ may a member of $\mathcal{M}^2_j$ and it becomes unclear how this node should participate: (a) As a member of set of nodes unreachable by $s_i$, $\overline{\mathcal{M}^1_i}$, or (b) As a member of set of nodes, $\mathcal{M}^2_j$, reachable by $s_j$. This needs to be clearly defined as the former event implies occurrence of failure in the network, while the latter does not.

  We take a different approach for the design of the protocol. For this consider the larger picture formed from executing the gossiping protocol $RG(2^i, 2^{i \cdot c})$ on the network. It may create many connected components $\mathcal{M}^1_i,\mathcal{M}^2_i,\dots$ etc. in the network, amongst which gossip can be realized. Messages may or may not be exchanged between nodes belonging to different components, but certainly nodes of every component need to know if there exist nodes outside their components, so that failure of the current phase is clearly identified and circulated in the network. Achieving this requires all the node of the network to transmit according to a certain schedule so that the following two requirements are simultaneously satisfied: While the nodes may be cooperating with other nodes in their own components to detect the failure, they should be also making their presence felt in other components at the same time, so that failure is inferred by the other components. We obtain the following result:

\begin{theorem}
\label{ack:gossiping}
  There exists a deterministic protocol that accomplishes acknowledged gossiping in all strongly connected networks of $n$ nodes, when nodes can take large labels and collision detection mechanism is available, with a round complexity of $O(n^2 \lg n)$.
\end{theorem}

\subsection{Selecting-Colliding Families}
\label{subsec:select-collide}
  The use of collision detection mechanism can be removed, if we can design a method to detect undiscovered neighborhood of a node. We realize that we may be able to use technology developed in the design of Select-And-Broadcast(.) protocol in \cite{KP04}, for this purpose. The properties of the transmission schedule which can enable a node to detect collision, with the help of discovered in-neighbor(s), is neatly abstracted as a combinatorial property of certain integral sets. It is referred to as {\it Selecting-Colliding} family. In particular, on following the transmissions scheduled according to Selecting-Colliding family, one of the two events must occur at every node, who has an undiscovered in-neighbor in the network: (a) Node receives the label of some undiscovered neighbor if it exists. (b) A collision occurs in a round, in which a single known neighbor and one or more unknown neighboring nodes transmit. The former case obviously identifies an undiscovered in-neighbor of a node. For the latter case, $\phi$ is received, which is what is received when no in-neighbors of the node transmits. Since a known in-neighbor of the node also transmitted in the same round, the node infers the existence of undiscovered neighborhood on receiving $\phi$. Following definition summarizes these requirements set theoretically:

\begin{definition}
\label{selecting-colliding}
  Let $N_{uk}$ and $N_{k}$ be sets for which it holds that $|N_k| < m$ and $|N_{uk}| < m^c$. A family $\mathcal{SSF}(m,m^c)$ of subsets $\{ S_1, S_2, \dots, S_r\}$ of $[1,\dots,m^c]$ is called a Selecting-Colliding family if $\exists S_i \in \mathcal{S_i}$ for which one the following two condition holds true:
\begin{itemize}
\item $|S_i \bigcap N_k| = 0$ and $|S_i \bigcap N_{uk}| = 1$.
\item $|S_i \bigcap N_k| = 1$ and $|S_i \bigcap N_{uk}| \geq 1$.
\end{itemize}
\end{definition}

  We prove the existence of Selecting-Colliding family $\mathcal{SSF}(n,n^c)$ of size $O(n^2 \lg n)$ using the probabilistic method.

\subsection{Acknowledged gossiping without collision detection mechanism}
\label{subsec:ackgpcd}
  Selecting-Colliding families are employed at the appropriate stage in the protocol for acknowledged gossiping, when collision detection mechanism is available, from Theorem \ref{ack:gossiping} to obtain the following.

\begin{theorem}
\label{ack:gossiping-nocd}
  There exists a deterministic protocol that accomplishes acknowledged gossiping on all strongly connected radio networks of $n$ nodes, when no collision detection mechanism is available and nodes can take large labels, in $O(n^2 \lg n)$ rounds.
\end{theorem}

\subsection{Fast acknowledged broadcasting in bidirectional networks}
  We adapt the ideas in \cite{V11}, to conduct a depth first search on bidirectional networks, to achieve fast acknowledged broadcasting when nodes can take large labels.

\begin{theorem}
\label{theo:bidir}
  There exists a deterministic protocol that accomplishes acknowledged broadcasting on all bidirectional radio networks of $n$ nodes, when nodes can take large labels, in $O(n \lg n)$ rounds.
\end{theorem}

\subsection{Acknowledged broadcasting for the collision model in \cite{BGIE}}
\label{subsec:ackbr}
  The following very weak model for handling collisions between radio transmissions was considered in \cite{BGI}: If two or more neighborers of a node transmit in the same round, then one of the two type of events can occur: (a) Message of one of the nodes is received (b) $\phi$ is received. It is easy to show that any Acknowledged protocol for this model cannot terminate in any specific number of rounds. The importance of this observation lies in emphasizing that efficient protocols for Acknowledged broadcasting/gossiping problem should somehow employ the technology for collision detection using helper nodes developed in \cite{KP04} (this technology is also used by protocols in \cite{UCW07}, \cite{FP08}).

\section{Acknowledged broadcasting on strongly connected networks}
\label{sec:arbllstsl}
  We present a deterministic protocol for acknowledged broadcasting on strongly connected networks, when nodes can have polynomially large labels. The protocol proceeds in phases $i = \{4,\dots,\lg n^c\}$. In phase $i$, nodes with labels less than $2^i+1$ are called {\it small}, with labels greater than $2^i+1$ but less than $2^{i \cdot c}+1$ called {\it medium} and with labels greater than $2^{i \cdot c}+1$ are called {\it large}.

\begin{figure}[h]
\begin{minipage}[t]{\textwidth}
\begin{algorithm}[H]

  Let $c, d$ be any pre-defined constants. Protocol proceeds in phases $i = \{4, 5, 6, \dots,\}$, where the $i^{th}$ phase consists of the following five stages:

\begin{center}

\begin{enumerate}
\item[] {\bf Stage 1:} Source node $s$ initiates the execution of broadcast protocol $RB(2^i, 2^{i \cdot c})$ on the network. Only small and medium nodes, while big nodes remain silent. All nodes who receive a message in this stage of the protocol are called {\it informed} and all those who do not receive a message in this stage are called {\it uninformed}.
\newline

\item[] {\bf Stage 2:} Only {\it small} and {\it medium} nodes that initiate or receive a message in Stage 1 of Phase $i$ participate, while others remain silent. Participating nodes execute deterministic gossiping protocol $RG(2^i, 2^{i \cdot c})$ twice: In the first execution, the nodes disperse their labels. In the second execution, they disperse the set of labels they received in the first execution.
\newline
  Let $\mathcal{M}_i$ denote the set of those nodes, who find that their labels belonged to the source message transmitted in the second execution of $RG(2^i,2^{i \cdot c})$.
\newline

\item[] {\bf Stage 3:} Nodes in $\overline{\mathcal{M}_i}$ transmit "failure" in every round of this stage. Furthermore, all nodes in $\mathcal{M}_i$ simulate gossip protocol $RG(2^i, 2^{i \cdot c})$ once more. By "simulation", it is meant that irrespective of whether or not a node receives a message in any given round or not, it still follows the same transmission schedule as that of the previous execution of this gossip protocol in Stage $2$.
  All nodes in $\mathcal{M}_i$ that do not receive any message in this Stage are set to a {\it failed} state.\\

\item[] {\bf Stage 4:} Nodes in $\mathcal{M}_i$ execute $RG(2^i, 2^{i \cdot c})$ once again where nodes, that were set to "failed" state in Stage $4$, propagate the "failure" message this time.\\

\item [] {\bf Stage 5:} If source $s$ was set to {\it failed} state, or receives a "failure" message in previous rounds, then the source learns that broadcast has failed. The "NACK" is dispersed by the source to the rest of the relevant network by execution of $RB(2^i, 2^{i \cdot c})$ and the protocol proceeds to the next phase;
  Otherwise, by way of silence nodes learn that broadcasting has completed and all nodes know it and know that other nodes know it and so on.
\end{enumerate}
\end{center}
\caption[Acknowledged broadcasting]{\label{prot:ackbst} Acknowledged broadcasting with large labels}
\end{algorithm}
\end{minipage}
\end{figure}

\begin{theorem}
\label{theo:ackbsc}
  Protocol \ref{prot:ackbst} accomplishes acknowledged broadcasting, on all strongly connected networks $N$, when nodes can be assigned large labels, in $O(NRG(n,n^c))$ rounds.
\end{theorem}

\begin{proof1}
  The proof of correctness is based on the following obvious lemma about the set $\mathcal{M}_i$ defined for Stage $2$ of Protocol \ref{prot:ackbst}.

\begin{lemma}
\label{lemma:arg1}
  A single execution of gossip protocol $RG(2^i,2^{i \cdot c})$ achieves gossip amongst the nodes in Set $\mathcal{M}_i$.
\end{lemma}

  We only need to show that if there are nodes in $\overline{\mathcal{M}_i}$, then a failure message is generated in the network, which reaches the Source before Step 5.

  If there are nodes in $\overline{\mathcal{M}_i}$, then they know it and are set to the {\it failed} state in Stage 4. We need to only make sure that stage 3 and stage 4 will convey any occurrence of failure anywhere in the network to the Source node. In the subsequent Stage, this failure is rebroadcasted via $RB(2^i, 2^{i \cdot c})$ and reaches all the relevant nodes in the network.

  Suppose that $\overline{\mathcal{M}_i} \neq \phi$ and broadcasting is not completed. All nodes in $\overline{\mathcal{M}}_i$ transmit in every round of Stage $3$. Furthermore, there must exist a path from the nodes in $\overline{\mathcal{M}_i}$ to the source node along which this message is transmitted and must transit from a node in $\overline{\mathcal{M}}_i$ to a node in $\mathcal{M}_i$ at some step. Thus, either "failure" is received by some node in $\mathcal{M}_i$, or there exists at least a node in $\mathcal{M}_i$ which hears only "silence" during the entire duration of Stage $4$ and hence set to the "failed" state at the termination of this Stage.

  In Stage $4$ of the protocol, "failure" is gossiped back to the source node, which is then rebroadcast-ed to the entire relevant network in Stage $5$ and the protocol proceeds to the next phase.

  Summing up the running time of the different stages over the different phases it is easy to verify that the running time of the above protocol is of the form $\sum_{i=1}^{i=l} (a \cdot NRG(2^i, 2^{i \cdot c}) + b \cdot NRB(2^i, 2^{i \cdot c}))$, for some constants $a$ and $b$. We need to only show that the protocol will conclude before $l = \lceil \lg_2 n \rceil + 1$ phases. To check this, note that the protocol may fail in a previous phase only if the number of nodes and the maximum size of label of a node does not fall in the range $[1,\dots,n]$ and $[1,\dots,n^c]$ respectively. In particular, when phase $i$ is executed with $i = \lfloor \lg_2 n \rfloor$, then $RG(n,n^c)$ and $RB(n,n^c)$ succeed in reaching all nodes and broadcast is completed and acknowledged via silence in future rounds.
\end{proof1}

\section{Acknowledged gossiping with collision detection mechanism}
\label{sec:gossip}

\begin{minipage}[t]{0.85 \textwidth}
\begin{algorithm}[H]
\begin{center}
  Initialize, $a=1$, $i=4$. Let $b=2^{c \cdot i - 1}$. Protocol proceeds in phases $i=4,3,4,\dots$. Each phase of the protocol consists of two segments:
\begin{enumerate}
\item[] {\bf First Segment} Nodes attempt to deduce if $RG(2^i, 2^{i \cdot c})$ can successfully achieve gossiping in the network. If not, then {\it failure} is generated somewhere in the network and circulated amongst all relevant nodes, who proceed to execute the next phase $i+1$ of the protocol; Else, the nodes successfully achieve gossiping in the second segment.

\begin{itemize}
\item[] {\bf Stage 1:} All small and medium nodes participate in this Stage. Nodes execute gossip protocol $RG(2^i, 2^{i \cdot c})$, by which they disperse their labels to the rest of the network. Each node $s$ compiles the list of labels $L_s^i$ of nodes from whom they receive message.\\

\item[] {\bf Stage 2:} Nodes execute gossip protocol $RG(2^i, 2^{i \cdot c})$ again, dispersing the corresponding sets of labels received by them in the previous Stage of this phase.

  By Lemma \ref{lemma:arg1} and Corollary \ref{gossip2}, after the second execution of $RG(2^i, 2^{i \cdot c})$, each node small and medium nodes $s$ learns the list $\mathcal{M}^i_s$ of nodes that it can reach and that can reach it in a single execution of $RG(2^i, 2^{i \cdot c})$.\\

\item[] {\bf Stage 3:} The purpose of this Stage is to detect failure and prepare for its dissemination.\\
  The following set of nodes initialize themselves to {\it failed} state: (a) All big nodes (b) All small and medium nodes for which $L_s^i \neq M_s^i$, i.e., nodes that can receive message from other nodes but cannot send them. (c) All small and medium nodes for which $|M^i_s| \geq 2^i$.\\
  All failed nodes transmit a failure message in every round of this Stage, along with their labels. All {\it unfailed} small/medium nodes transmit their labels according to Strongly selective family $SSF(2^i+1, 2^{i \cdot c})$.\\
  Following nodes are also set to failed state after this execution: (A) Any node that receives a failure message (B) Any node $s$ that {\it detects collision} in a round in which at most one node from list $M_s^i$ from Stage $2$ transmitted. (C) Any node $s$ that receives a label $l$ of some node which does not belong to list $M_s^i$ constituted in Stage 2.

\item[] {\bf Stage 4:} In this Stage, a (potentially) small subset of failed nodes spread failure to all relevant network.
  For this, Gossip protocol $RG(2^i, 2^{i \cdot c})$ is executed in which all small and medium nodes set to failed state in Stage $3$, disperse the "failure" message.
  Either all small and medium nodes receive "failure" message in Stage $4$, or were set to failed state in Stage $2$ or Stage $3$, and learn that the protocol should proceed to the next phase, i.e., $i=i+1$;\\
  Or, none of the nodes generate "failure" message and all nodes learn of successfull completion of this segment of the protocol.
\end{itemize}
\item[] {\bf Second Segment:} Nodes execute this segment only if the first segment is completed successfully. It lasts for $NRG(2^i, 2^{i \cdot c})$ rounds, in which nodes execute the gossip protocol $RG(2^i,2^{i \cdot c})$ to disperse their messages in the network.
  Otherwise, it consists only of empty rounds when the nodes wait to execute the next phase.
\end{enumerate}
\end{center}
\caption[Acknowledged Gossiping]{Acknowledged gossiping with collision detection mechanism\label{prot:ackgpcd}}
\end{algorithm}
\vspace{-0.25in}
\end{minipage}

    Protocol \ref{prot:ackbst} for acknowledged broadcasting can also clearly achieve gossiping amongst all the nodes. However, the reason it is not a protocol for acknowledged gossiping is because there is no pre-defined "leader" at the start of a gossiping protocol, who plays the role of the leader in the broadcast procedure. Thus, it seems sufficient to design a protocol that elects a leader. For the reasons elaborated in Subsection \ref{subsec:ackgpnocd}, we take a different approach: Instead of designing a protocol for binary Selection, we use the gossip protocol $RG(2^i,2^{i \cdot c})$ multiple times to achieve our goal.

  Our protocol has the same phase-wise structure as other protocols for the acknowledged problems. The reader may review the nomenclature in Subsection \ref{notations}, which is followed in this Section also. The design of Protocol \ref{prot:ackgpcd} is based on the following observation deduced along the same lines as Lemma \ref{lemma:arg1} for Protocol \ref{prot:ackbst}.

\begin{corollary}
\label{gossip2}
  Let gossip protocol $RG(2^i, 2^{i \cdot c})$ be executed twice on any network $N$ (of arbitrary size), where in the first execution nodes simply disperse their labels. In the second execution, nodes disperse the list of all labels received by them in the first execution. Then, after the second execution, each small and medium node $s$, can compile a list of nodes $\mathcal{M}^i_s$ that they can reach and who can reach them in a single execution of gossip protocol $RG(2^i, 2^{i \cdot c})$.

  If $\mathcal{M}^i_s$ doesn't include the labels of all the nodes in the entire network or consists only of a single node, then $n > 2^i$ and gossiping is not completed.
\end{corollary}


\begin{theorem}
\label{gossip}
  Protocol \ref{prot:ackgpcd} achieves Acknowledged gossiping on all strongly connected networks of $n$ nodes in $O(n^2 \lg n)$ rounds, when nodes can take polynomially large labels.
\end{theorem}

\begin{proof1}
  If Protocol \ref{prot:ackgpcd} runs till $i = \lceil \lg n \rceil^{th}$ phase, then gossiping is accomplished successfully in the Second Segment of the last phase (follows from correctness of vanilla gossip protocol $RG(.,.)$). We only need to show that, if gossiping is not successfully completed in $j^{th}$ phase for $j < \lceil \lg n \rceil$, then the next phase of the protocol is executed.

  For every node $s$ a list $M^i_s$ is prepared in Stage $2$. If $L^i_s \neq M_s^i$, then node $s$ detects failure in Stage $2$. Every node $v$ in $M_s^i$ also detects failure in stage $4$.

  Let $M_u^i$ be a list, corresponding to some node $u$, generated in stage $2$ of phase $i$ of Protocol \ref{prot:ackgpcd}, for which failure is not detected by any node of component $M_u^i$, up till stage $2$. If failure is detected by any node of $M_u^i$, then it is gossiped in the entire component in stage $4$. We will show that some node $v$ in $M_u^i$ will detect failure in stage $3$ of phase $i$ of Protocol \ref{prot:ackgpcd} and this failure will reach all nodes in list $M_u^i$ in the next stage.

  Since the network is strongly connected, there must exist at least one node $v$ in $M_u^i$, which has an in-neighbor in $\overline{M_u^i}$. If $M_u^i$ consists of a single node, or if $|M_u^i| > 2^i-1$, then $u$ is set to to the fail state at the beginning of Stage $3$. So, we may assume that $|M_i^u| \leq 2^i-1$.

  Let set $S_v$ denote the unknown in-neighbor of node $v$, which lies outside $M^u_i$ and $S_v'$ denote the neighborhood of $v$ that belongs to $M^u_i$. We have the following two possibilities:
\begin{enumerate}
\item{} If there is any large node $l$ in $S_v$, then it transmits failure in every round of Stage $3$. When the transmission schedule according to strongly selective family $SSF(2^i,2^{i \cdot c})$ is executed, then by the Definition \ref{stronglyselectivefamily} of Strongly Selective Family $SSF(2^i,2^{i \cdot c})$, there exists at least one subset $T \in SSF(2^i,2^{i \cdot c})$ for which $|T \cap S_v'|=1$ (in fact, this holds for many subsets of the SSF). In this particular round, corresonding to subset $T$, exactly one node from $S_v'$ transmits. Also, since the large node $l$ transmits in every round of Stage $3$, a collision occurs, which is detected by the collision detection mechanism. Since node $v$ knows that at most one node from its known neighborhood $S_v'$ has transmitted, it infers that collision must involve transmission from at least one more node which does not belong to $S_v'$. $v$ infers that $S_v \neq \phi$ and sets itself to failure.

\item{} Else, all in-neighbors of node $v$, belonging to set $S_v$, are only small and medium nodes. We also know that $|S_v'| \leq 2^i-1$. Let $R$ be any subset obtained by including all elements of subset $S_v'$ and one element $z$ of $S_v$. By Definition \ref{stronglyselectivefamily}, there must exist a subset $S_r \in SSF(2^i,2^{i \cdot c})$ for which $S_r \bigcap R = {z}$. That is, there exists a round in which none of the known in-neighbors of $v$ and at least one unknown in-neighbor $z$ of $v$ transmit. When the transmission schedule corresponding to the Strongly selective family $SSF(2^i,2^{i \cdot c})$, is executed, either label of $z$ is received by node $v$, or a collision happens and is detected by $v$ using the collison detection mechanism. Since $v$ knows that none of its known in-neighbors $S_v'$ transmitted in the round, it infers that collision must have occured due to transmissions from undiscovered neighbors of $v$. In either case, $v$ discovers the existence of undiscovered neighbors and initializes itself to failure.
\end{enumerate}

  Thus, there exists at least one node $v$ in every list/component $M_u^i$ which detects failure if the gossip protocol $RG(2^i,2^{i \cdot c})$ is not successfully executed in Stage $3$. This failure is dispersed throughout that component/list in the next stage $4$ and all nodes in the network know that gossip has not successfully completed. Nodes of the network proceed to the next phase of the protocol.
  Nodes in the network finally execute the second segment of the protocol \ref{prot:ackgpcd} to disperse their messages when successfull termination of gossip happens in the previous segment of a phase of the protocol.

  The number of rounds that Protocol \ref{prot:ackgpcd} takes in $i^{th}$ phase is $O(NRG(2^i, 2^{i \cdot c}) + SSF(2^i,2^{i \cdot c}))$. Also, it follows from the above proof of correctness that protocol does not run beyond $\lceil \lg n \rceil$ phase. If the protocol runs for $i = \lceil \lg n \rceil$ phases, then the total number of rounds taken by it is $\sum_{i=1}^{i=\lceil \lg_2 n \rceil} O(NRG(2^i, 2^{i \cdot c}) + SSF(2^i,2^{i \cdot c})) = O(n^2 \lg n)$ rounds.
\end{proof1}

\begin{remark}
  Since the round complexity of the protocol is governed by the size of Strongly Selective family $SSF(n,n^c)$, we have not concerned with optimizing the number of invocations of protocol $RG(2^i, 2^{i \cdot c})$ in our protocol.
\end{remark}

\section{The Selecting-Colliding families}
\label{selectingcolliding}
  Protocol \ref{prot:ackgpcd} uses the collision detection mechanism. We precisely capture the essential combinatorial requirements for the transmission schedule in Stage $3$, so that nodes can discover existence of undiscovered in-neighbors without collision detection mechanism. The collision inference is achieved by employing the technological innovation in \cite{KP04}, in design of Echo() and Select-And-BroadCast() protocol, which is based on Clause (4) of the model described in Section \ref{sec:model-and-tools}. In set theoretic terms, the transmission schedule of nodes of the network may be seen as a family of sets, which is referred to as {\it Selecting-Colliding} family. We shall present a formal definition of {\it Selecting-Colliding family} and show its existence using the probabilistic method.

  Let $N^u_k$ and $N^u_{uk}$ refer to the subset of known and unknown small, medium in-neighbors of a node $u$ of the network. Furthermore, it is also known that $|N^u_k| < l$. The execution of transmission schedule outlined by Selecting-Colliding family should result in all nodes in the network to learn about the existence of undiscovered in-neighbors, as long as they have certain restrictions on the size of their known in-neighborhood. The following definition captures these requirements.

\begin{definition}
\label{selectingcolliding-family}
  A family of subsets $SCF(l,l^c)$ is a Selecting-Colliding Family for every subset $N^u_k \subset \{1,2,\dots,l^c\}$, for which $|N^u_k| < l$, if for every $N^u_{uk} \subset \{1,2,\dots,l^c\}$, $N^u_{uk} \neq \phi$, there exists a subset $S_i$, $S_i \subset \{1,2,\dots,l^c\}$, $S_i \in SCF(l,l^c)$, for which (at least) one of the following two conditions hold true:
\begin{itemize}
\item[] {\bf Condition A:} $|S_i \bigcap N^u_{uk}| = 1$ and $|S_i \bigcap N^u_k| = 0$.
\item[] {\bf Condition B:} $|S_i \bigcap N^u_{uk}| \geq 1$ and $|S_i \bigcap N^u_k| = 1$.
\end{itemize}
\end{definition}

\vspace{0.1cm}
\begin{theorem}
\label{theo:sc}
  Let $l$ be positive integer $>3$. There exists a Selecting-Colliding family, Definition \ref{selectingcolliding-family}, of size $O(l^2 \cdot \lg l)$.
\end{theorem}
\vspace{-0.5cm}
\begin{proof1}
  We demonstrate the existence of $SCF(l,l^c)$ using the probabilistic method. Our proof only targets the (Condition B) of Definition \ref{selectingcolliding-family}. That is, we show that there exists a family of subsets of $\{1,2,\dots,l^c\}$, $SCF(l,l^c)$ such that $|SCF(l,l^c)| \in O(l^2 \lg l)$, such that $\forall N^u_k = |N^u_k| < l$, $\forall N^u_{uk}: |N^u_{uk}| \neq \phi$, there exists $i \in [1,\dots,l]$ for which the following two conditions hold true: \begin{enumerate} \item{} $|S_i \bigcap N^u_{uk}| \geq 1$.  \item{} $|S_i \bigcap N^u_k| = 1$.
\end{enumerate}

  The family $SCF(l,l^c)$ is obtained by taking union of two collections, $SCF_1$ and $SCF_2$, where $SCF_1$ targets the case when $|N^u_{uk}| \geq l$ and $SCF_2$, the case when $|N^u_{uk}| \le l$.
\newline

\begin{enumerate}
\item{\bf $|N^u_{uk}| \geq l$}\\
\label{greaterthan}
  We show that a family of subsets, each of whose members is chosen with appropriate probability values, achieves the desired outcome with non-zero probability. The family of subsets chosen by us will satisfy condition B of Definition \ref{selectingcolliding-family}.

  We build $SCF_1$ by taking the union of smaller families which cater to different sizes of set $N_k^u$.
  Let $|N^u_k| = x$ and $|N^u_{uk}| = l$. First, we choose members of $SCF_1$ for those $x$, for which $\frac{m}{2} \leq x \leq m$ of $x$. We call these members $SCF_1^m$. Then, $SCF_1 = \bigcup_{i=0}^{i=\lceil \lg l \rceil} SCF_1^{2^i}$.

  Let $S_j$ be a set, obtained by independently choosing each element from $\{1, 2, \dots, l^c\}$ with probability $b = \frac{1}{m}$. Let $p_i(x)$ denote the probability that $S_j$ hits $N^u_k$ exactly once and $N^u_{uk}$ at least once. Then,
\[
  p_j(x) = [x * b * (1 - b)^{x - 1}] * [1 - (1-b)^{l}]
\]
  where $b = \frac{1}{m}$, and the first term refers to the probability that $S_i$ hits $N^u_k$ exactly once and the second term refers to the probability that $S_i$ hits $N^u_{uk}$ at least once. That is,

\[
  p_j(x) = [x * \frac{1}{m} * (1 - \frac{1}{m})^{x - 1}] * [1 - (1-\frac{1}{m})^l]
\]

  Noting that $(1 - \frac{1}{m})^m < e^{-1}$ and so $1 - e^{-a} < (1 - \frac{1}{m})^{m \dot a}$ for all positive $a$, we have that
\[
  p_j(x) = [x * \frac{1}{m} * (1 - \frac{1}{m})^{x - 1}] * [1 - (1-\frac{1}{m})^l]
     \geq [\frac{x}{m} * (1 - \frac{x-1}{m})] * [1 - e^{-\frac{l}{m}}]
\]

  For $\frac{m}{2} \leq x < m$, the term $\frac{x}{m} - \frac{x^2-x}{m^2}$ is minimized at $x = m$. At $x = m$, $p_j(x)$ simplifies to:
\[
    \geq [\frac{m}{m} \cdot \frac{1}{m} \cdot (1 - \frac{m-1}{m})] \cdot [1 - e^{- \frac{l}{m}}]
\]

\[
    = [1 \cdot \frac{1}{m}] \cdot [1 - e^{-\frac{l}{m}}].
\]

\[
    = [\frac{1}{m} \cdot [1 - e^{-\frac{l}{m}}].
\]

  Also, $e^{\frac{-l}{m}} > e^{-1}$ as $l > m$.

\[
  p_j \geq \frac{1}{m} * (1 - \frac{1}{e}) > \frac{1}{2 \cdot m}.
\]

  $1-p_j$ is the probability that a randomly chosen set $S_j$ does not {\it Hit $N_k^u$ once and $N^u_{uk}$ at least once}. Let $SCF_1^m$ consist of $t$ such sets chosen in the same manner. The probability that all the $t$ sets, chosen in the above manner, fail for some $N^u_k$ whose size lies in range $[\frac{m}{2},m]$ and $N^u_{uk}$ of size $f(l)=l$ is the probability of failure.

  Let $p_f = Pr [ SCF_1^m$ does not satisfy condition B of Definition \ref{selectingcolliding-family} $]$, for $\frac{m}{2} \leq |N_k^u| \leq m$. By union bound we have:
\[
  p_f \leq \sum_{x=\frac{m}{2}}^{x=m} \binom{l^c}{x} \cdot \binom{l^c}{l} \cdot (1-p_j)^t.
\]

  Note that, $\binom{l^c}{l}$ is the number of sets $N_{uk}^u$ of size $l$. Since, we intend to hit $N_{uk}^u$ at least once, it is sufficient to just ensure that each of the $\binom{l^c}{l}$ sets is hit by some set $S_j$. If this is ensured for all sets of size $l$ then this property is ensured for all supersets of these sets. The failure probability may be upper bounded by looking at only these $\binom{l^c}{l}$ sets.

  Using the fact that $\binom{a}{b} \le \left ( \frac{e \cdot a}{b} \right )^b$, we have
\[
  p_f \leq \sum_{x=\frac{m}{2}}^{x=m} \left ( \frac{e \cdot l^c}{x} \right )^{x} \cdot \left ( \frac{l^c \cdot e}{l} \right )^{l} \cdot (1 - \frac{1}{2 \cdot m})^t
\]

  For different values of $x \in [\frac{m}{2},m]$, the term in the summation is maximized at $x = m$. Thus,
\[
  p_f \leq  m^2 \cdot \left ( \frac{e \cdot l^c}{\frac{m}{2}} \right )^{\frac{m}{2}} \cdot \left (\frac{l^c \cdot e}{l} \right )^{l} \cdot (1-\frac{1}{2 \cdot m})^t
\]

  We would like that $p_f < 1$. That is,
 \[
  m^2 \cdot \left ( \frac{e \cdot l^c}{\frac{m}{2}} \right )^{\frac{m}{2}} \cdot \left (\frac{l^c \cdot e}{l} \right )^{l} 
  < \frac{1}{(1-\frac{1}{2 \cdot m})^t}.
\]

  This inequality holds true for $t > d \cdot l \cdot m \cdot \lg l$, for suitably chosen constant $d$, since $l > m$. Thus, with a non-zero probability the random subsets in $SCF_1^m$ satisfy the requirements.

  Let $SCF_1 = \bigcup_{i=1}^{i=\lfloor \lg l \rfloor + 1} SCF_1^{2^i}$. Then, $SCF_1$ satisfies Definition \ref{selectingcolliding-family} for all $|N^u_k| < l$ and $|N^u_{uk}| \geq l$. Furthermore, $|SCF_1| = \sum_{i=1}^{i = \lfloor \lg l \rfloor + 1} |SCF_1^{2^i}| = \sum_{i=1}^{i = \lfloor \lg l \rfloor + 1} d \cdot l \cdot 2^i \lg l \leq  4 \cdot d \cdot l^2 \cdot \lg l$.

\item{\bf $|N^u_{uk}| < l$}\\
\label{lessthan}
  If $|N^u_{uk}| < l$, then the size of the total in-neighbor of a node is $\leq 2 \cdot l$, as it is given that $|N^u_k| < l$. For this case, we choose $SCF_2$ to be the Strongly Selective Family $SSF(2 \cdot l, l^c)$. All nodes receive the labels of their entire in-neighbors including nodes in $N_k^u, N^u_{uk}$ and hence discover any undiscovered in-neighbor.
\end{enumerate}
\vspace{0.25cm}

  Let $SCF(l,l^c)$ be the union of set of subsets $SCF_1$ from \ref{greaterthan} and set of subsets $SCF_2$ from \ref{lessthan}. Then, this family satisfies the Definition \ref{selectingcolliding-family}. The size of $SCF_1$ is $O(l^2 \cdot \lg l)$ and the size of $SCF_2$ is $O(l^2 \lg l)$, so the size of $SCF(l,l^c)$ is $O(l^2 \lg l)$.
\end{proof1}

\begin{remark}
\begin{enumerate}
\item{}
When $|N^u_{uk}| >> l$ and $|N^u_k| << l$, then sampling at probability $\frac{1}{l}$, which works for higher values $|N^u_k|$ is not useful. Such a sample is likely to exclude every member of $N^u_k$ and include at least two members of $N^u_{uk}$ and hence node $u$ is very likely to receive a collision which is indistinguishable from silence. On the basis of only this type of information, $u$ cannot distinguish case $|N^u_{uk}| = 0$ from case $|N^u_{uk}| > 1$.
\item{} When $|N^u_{uk}| < l$ and in particular only a constant, while $|N^u_k| = O(l)$, sampling at a high rate, say inverse of a constant, may successfully isolate a single element from $N^u_k$ but will is also likely to select multiple elements from $N^u_k$. This results in a collision in which multiple elements of $N^u_k$ are involved and hence receiving node $u$ cannot gather any information about $N^u_{uk}$, i.e., determine whether $|N^u_{uk}| \neq \phi$. On the other hand, if sampling is done at a low rate, say $O(\frac{1}{l})$, then a random subset has very low chance of capturing even a single element of $N^u_{uk}$.
\end{enumerate}
\end{remark}

\section{Acknowledged gossiping without collision detection mechanism}
\label{sec:NoCdProt}
  The definition of Selecting-Colliding families does not capture all possibilities handled in stage $4$ of Protocol \ref{prot:ackgpcd}. But it is sufficient to handle some extraneous conditions that are not explicitly captured by the requirements of $SCF$, Definition \ref{selectingcolliding-family}. For example, it is possible that $|N^u_{uk}| = \phi$, but there is a big node which is in-neighbor of node $u$. The presence of even large in-neighbor(s) is detected by the Selecting-Colliding family demonstrated in previous section. Recall that, in protocol \ref{prot:ackgpcd}, every large in-neighbor of $u$ transmits failure in every round of the protocol. By Condition (B) of definition of SCF($l,l^c$), the only condition exploited in the proof (of existence of SCF), there exists a subset $S_i$ for which $|S_i \bigcap N^u_k| = 1$. That is, exactly one known in-neighbor of $u$ transmits in the round corresponding to $S_i$. Since, at least two nodes transmit in the same round as a result, collision occurs and node $u$ received $\phi$. However, since $u$ knows that exactly one node in $N_k^u$ transmitted, some other node, in $\overline{N^u_k}$, should have also transmitted for $u$ to have received $\phi$ (Clause (4) of model described in Section \ref{sec:model-and-tools}).

  Protocol for Acknowledged gossiping without collision detection mechanism is obtained by replacing the Strongly selective family $SSF(2^i+1, 2^{i \cdot c})$ used in stage 3 of Protocol \ref{prot:ackgpcd}, by Selecting-Colliding family $SCF(2^i,2^{i \cdot c})$. For the sake of completeness, we present the full protocol below.

\begin{minipage}[t]{6.0in}
\begin{algorithm}[H]
\begin{center}
  Initialize, $a=1$, $b=2^{c \cdot i - 1}$. Protocol proceeds in phases $i=2,3,4,\dots$. Each phase of the protocol consists of two segments:
\begin{enumerate}
\item[] {\bf First Segment} Nodes try to conclude if $RG(2^i, 2^{i \cdot c})$ can successfully achieve gossiping in the network. If not, then {\it failure} is generated somewhere in the network, circulated amongst the nodes who proceed to execute the next phase $i+1$ of the protocol; Else, the nodes execute the second segment.

\begin{itemize}
\item[] {\bf Stage 1:} All nodes with labels in the range $[a,\dots,b]$ participate in this Stage. Nodes execute gossiping protocol $RG(2^i, 2^{i \cdot c})$, by which they disperse their labels to the rest of the network.

\item[] {\bf Stage 2:} Nodes execute the gossiping protocol $RG(2^i, 2^{i \cdot c})$ one more time. In this execution, nodes disperse their corresponding sets of labels received by them in the previous Stage of this phase.

  By Lemma \ref{lemma:arg1} and Corollary \ref{gossip2}, after the second execution of $RG(2^i, 2^{i \cdot c})$, each node $s$ learns the list $\mathcal{M}^i_s$ of nodes that it can reach and that can reach it in a single execution of $RG(2^i, 2^{i \cdot c})$.

\item[] {\bf Stage 3:}The following set of nodes initialize themselves to {\it failed} state: (a) All big nodes (b) All small and medium nodes whose list from Stage $2$ does not consist of any node from whom they received message in Stage $1$. (c) All small and medium nodes whose list from Stage $2$ is of size greater than $2^i$.\\
  All failed node transmit a failure message in every round of this Stage, along with their labels.\\
  All unfailed small and medium nodes transmit, their labels, according to schedule outlined by Selecting-Colliding Family $SCF(2^i, 2^{i \cdot c})$.\\
  The following nodes are also set to to failed state after this execution: (A) Any node that receives a failure message (B) Any node that detects {\it collision} in a round in which at most one node from its list of reachable nodes from Stage $2$ transmitted. (C) Any node that receives a label of node which does not belong to its list constituted in Stage 2.\\
  Either all small and medium nodes receive "failure" message or were set to failed state in Stage 2, Stage 3 or Stage 4, and know that the protocol should proceed to the next phase, i.e., $i=i+1$; Or, none of the nodes generate "failure" message and hence all nodes know of the completion of this segment of the protocol.\\
\end{itemize}

\item[] {\bf Second Segment} Nodes execute this segment only if no failure is generated in the first segment. It lasts for $NRG(2^i, 2^{i \cdot c})$ rounds, in which nodes execute the gossip protocol $RG(2^i,2^{i \cdot c})$ to disperse their messages in the network.
  Otherwise, it consists only of empty rounds when the nodes wait to execute the next phase.
\end{enumerate}

\end{center}
\caption[Acknowledged Gossiping]{Acknowledged Gossiping with Collision Detection Mechanism\label{prot:ackgpnocd}}
\end{algorithm}
\end{minipage}

  The proof of correctness of Protocol \ref{prot:ackgpnocd} follows the same line of argument as the proof of Theorem \ref{gossip}.

\begin{theorem}
\label{nocollide:gossip}
  There exists a deterministic protocol that accomplishes gossiping on all strongly connected networks of size $n$, when nodes take labels in range $[1,\dots,n^c]$, in $NRG(n, n^c)$ rounds, when no collision detection mechanism is available.
\end{theorem}


\section{Acknowledged broadcasting on bidirectional networks}
\label{sec:arbstbl}
  We present an efficient deterministic broadcast protocol that takes $O(n \cdot \lg n)$ rounds on all bidirectional networks of $n$ nodes, when nodes can have polynomially large labels.

  The main idea is as follows. The source first discovers a neighboring node in the network, using selective families for increasing sized subsets. This stage is called DISCOVER-FIRST-NEIGHBOR. Once a single neighboring node of source has been discovered in some round, the entire network is explored using a token in a DFS manner, using binary selection to select an undiscovered neighbor. The second stage is called DISCOVER-AND-PASS-TOKEN. This protocol uses ideas from \cite{KP04}, in which a protocol of round complexity $O(n \lg^2 n \lg \lg n)$ is given for bidirectional networks for deterministic gossiping. In the next subsection, we present adaptations of canonical tools which were introduced in \cite{KP04}. In subsection \ref{subsec:arbstbl}, we present the complete protocol for bidirectional networks.

\subsection{Some primitives for bidirectional networks}
\label{sec:discover}
  In this Section we present sub-protocols which are used for composing larger protocols. The first procedure, described in Subsection \ref{echo}, is used to estimate the size of set $A$ and is obtained by adapting the Binary-Selection-Broadcast() procedure in \cite{KP04}. It determines if the number of nodes in subset $A$ connected to a node $s$ is $0$, $1$ or more. This is achieved with the assistance of a helper node, whose label is $h$, where the helper node can be any node that the initiator of Estimate() protocol is aware of. The second procedure in subsequent Subsection \ref{binaryselectionbroadcast}, is used the first procedure to select a single undiscovered neighbor of a node.

\subsubsection{Estimating the size of a neighborhood}
\label{echo}
  Procedure Estimate() is initiated by a node $s$. It works with three parameters: (1) $X$ which is a subset of labels to be excluded from the search (2) $h$ which is the label of a particular node (3) $Y$ a subset of labels within which the search is to be carried out.

\begin{minipage}[t]{6.0in}
\begin{algorithm}[H]
\begin{center}

\begin{enumerate}
\item{} Node $s$ transmits message $(h, X, Y)$, which is received by all its neighbors in network $N$.

\item{} Every neighbor of $s$ with labels in subset $(Y - X) - \{h\}$ transmits its label.

\item{} All neighbors of $s$ in $(Y - X) - \{h\}$ transmit their labels. There are three possible effects of executing the above Steps.
\begin{enumerate}
\item{} A message is received in Step 2 and no message is received in Step 3. In this case, the node $s$ knows that it has a single neighbor with label in $(Y - X) - \{h\}$ and learns the label of this unique node.
\item{} No message is received in Step 2 but a message from $h$ is received in Step 3. In this case, the node $s$ knows that it has no neighbors with labels in $(Y - X) - \{h\}$.
\item{} No message is received in Step 2 or Step 3. In this case, the node $s$ concludes that $s$ has at least $2$ neighbors in $(Y - X) - \{h\}$.
\end{enumerate}
\end{enumerate}
\end{center}

\caption{Estimate($h$, $X$, $Y$)\label{prot:estimate}}
\end{algorithm}
\end{minipage}


\subsubsection{Binary selection of an undiscovered node}
\label{binaryselectionbroadcast}
  The goal of this procedure is for a node $s$, to discover a new node in its undiscovered neighborhood $\overline{X}$, with the help of an already discovered assistant node $h$.

\begin{minipage}[t]{6.0in}
\begin{algorithm}[H]
\begin{center}

\begin{enumerate}
\item{} Node $s$ transmits the label of a neighboring node $h$, which is called helper of $s$ and the subset $X$ of labels of already discovered neighbors.

\item{} $s$ first determines if there exists an undiscovered node in its neighborhood by executing Estimate($h$, $X$, $[1,\dots,n^c] - X$). If the entire neighborhood of $s$ is discovered, then the procedure Binary-Select() terminates, else it continues to the next step.

\item{}
\begin{enumerate}
\item{} Initialize $i$ so that: $2^{i-1} \leq$ (maximum value of a label in $X$) $\leq 2^i$.
\item{} $s$ initiates execution of Estimate($h$, $X$, $2^i$).
\item{} If a single new neighbor is discovered, then its label $l$ is learned by $s$ and the procedure terminates;\\
       Else, if it is found that the number of undiscovered neighbors of $s$ is $\geq 2$, then this step terminates and continues to step $4$.\\
       Else, $i = i+1$ and sub-steps (b) and (c) of this step are again executed with incremented value of $i$.\\
\end{enumerate}

  At termination of step $3$, nodes conclude that at least $2$ undiscovered neighbors of $s$ have labels in the range $[1,\dots,2^i]$ and continue the search in this range.\\

  Each of the remaining steps $4, 5, 6, \dots$ consists of the following binary search stage: Nodes execute Procedure Estimate($h$, $X$, $Y$), with range $Y = [a=1,\dots,b=2^{i-1}]$), initiated by $s$. If the answer received from the execution is the label of a single undiscovered neighbor of $s$, then the procedure terminates and the value of this label is returned; If the answer received from the execution is $0$, then in the next step nodes continue the search the complimentary range $Y = [a=2^{i-1}+1,\dots,b=2^i]$, else the nodes continue the search on range $Y = [a=1,\dots,b=2^{i-2}]$, and so on and so forth.
\end{enumerate}

\end{center}
\caption{Binary-Select($s,h,X$)\label{binary-select}}
\end{algorithm}
\end{minipage}

\subsection{Acknowledged broadcasting on bidirectional networks}
\label{subsec:arbstbl}
  We present a deterministic broadcast protocol that takes $O(n \cdot \lg n)$ rounds, when nodes can have polynomially large labels.

  The main idea is as follows. The source first discovers a neighboring node in the network, using selective families for increasing sized subsets. This stage is called DISCOVER-FIRST-NEIGHBOR. Once a single neighboring node of source has been discovered in some round, the entire network is explored using a token in a DFS manner, using binary selection to select an undiscovered neighbor. The second stage is called DISCOVER-AND-PASS-TOKEN. This protocol uses ideas from \cite{KP04}, in which a protocol of round complexity $O(n \lg^2 n \lg \lg n)$ is given for bidirectional networks for deterministic gossiping.

\begin{minipage}[l]{6.0in}
\begin{algorithm}[H]
\begin{center}
  Let $c, d$ be any pre-defined constants. Protocol proceeds in phases $i = \{4, 5, \dots,\}$, where the $i^{th}$ phase consists of the following two stages:

\begin{enumerate}
\item[] {\bf DISCOVER-FIRST-NEIGHBOR:}
  If the source has already designated a {\it helper} node in the previous phase, then all nodes wait out till the start of the next stage. Else they proceed to discover a neighbor of the source node as below.\\

  In the $i^{th}$ iteration, all nodes which have received a message in round $0$, execute ($2^i,(2^i)^c$)-selective family \ref{theo:selectivefamily}, to schedule their transmissions, of their label values. In odd rounds, transmissions chosen from the appropriate set of selective family takes place (nodes just transmit their labels), and in even rounds the sources echoes and re-transmits the label of the node from which it received the message, if it did.\\
  If a neighbor $h$ of source $sd$ is discovered, then all neighbors of source learn its label in the next round. The discovered neighbor, with label $h$, is marked as a helper node of source $s$.\\

  Once a helper node has been identified in round $i$, source and its neighbors wait till termination of this stage i.e., round $2 \cdot SF(2^i,(2^i)^c)$.

\item[] {\bf DISCOVER-AND-PASS-TOKEN:}
  This stage is executed, if the Source node has been able to identify a helper node in the previous Stage and lasts for $7 \cdot (2^i \lg 2^i)$ rounds, irrespective of that.\\
   If an initial helper node $h$ was successfully identified in the previous Stage, then the source initiates the broadcasting in this round. Otherwise, all nodes just wait (do nothing) for the Stage to terminate, which automatically happens after $5. 2^i \lg 2^i$ rounds. The set of already discovered nodes is maintained in a subset $X$, which is passed along with a TOKEN from a node to node in a DFS manner.\\

\begin{enumerate}
\item[] {\bf First Step} The subset is $X$ is initialized as follows: $X = \{ s \}$. Since neighbor $h$ of $s$ was discovered in previous Step, the token is passed to node $h$, in this Step.

\item[] {\bf $i^{th}$ Step} Let $r$ be the node which received the token in the previous Step, from some node $h$. Then, subset $X$ is first incremented: $X = X \bigcup \{ r \}$ and $r$ transmits the message $m$ to be broadcasted. Using the helper node $h$, node $r$ attempts to discover a new neighbor by execution of Binary-Select($r,h,X$). The result of this can be one the following:
\begin{enumerate}
\item If no new node is discovered, then the token is returned back to the node $p$ from which node $r$ had received the token for the first time.

\item If a new node is discovered, so that the total number of discovered nodes $|X|$ is now $2^i+1$, then further action is abandoned and broadcasting is restarted in the next phase $i+1$.

\item If a new node $t$ is discovered, so that total number of discovered nodes $|X|$ is less than $2^i$, then the token is passed to $t$. The node $r$ from which the token is passed is identified as a helper to $t$ and this step is executed again.\\
\end{enumerate}
\end{enumerate}

  Source waits for token to reach it along with set $X$. If the token reaches the source before round $4 \cdot 2^i \lg 2^{c \cdot i}$ and it checks that none of its neighbors are unexplored, then the source knows that broadcast has been completed. To reconfirm it, source re-broadcasts this acknowledgment to the entire network in the last $3 \cdot 2^i \lg 2^{c \cdot i}$ rounds.\\

  If nodes receive the reconfirmation before the termination of this phase, then they know that broadcast has been completed and acknowledged and the protocol terminates.\\

  Else, the protocol proceeds to the $(i+1)^{th}$ phase and nodes know of it by hearing silence in the last $3 \cdot 2^i \lg 2^{c \cdot i}$ rounds.
\end{enumerate}
\end{center}
\caption{Acknowledged broadcasting on bidirectional networks \label{alg4arbllstbl}}
\end{algorithm}
\end{minipage}

\begin{theorem}
\label{arb-ll-st-bl}
  Protocol \ref{alg4arbllstbl} achieves acknowledged broadcast on all bidirectional networks with $n$ nodes in $O(n \lg n)$ rounds, when nodes can take polynomially large labels.
\end{theorem}
\begin{proof1}
  The correctness of the protocol follows from correctness of DFS procedure, definition of selective families, correctness of procedure Estimate(), procedure Binary-Select() given in Section \ref{sec:discover}, adapted from \cite{KP04}.

  The number of rounds taken by Protocol \ref{alg4arbllstbl} on bidirectional networks of $n$ nodes, in the worst case is: $\sum_{i=2}^{i \leq \lg n} (SF(2^i,2^{i \cdot c}) + 7 \cdot i \cdot \lg i^c$, where $SF(k,n)$ is the minimum size of a $(k,n)$-selective family. From Definition \ref{selectivefamily}, we have that $SF(k,n) \in O(k \lg \frac{n}{k})$. Adding up, the round complexity of the broadcast protocol is $O(n \lg n)$ rounds.
\end{proof1}

\section*{Acknowledgements} I am very grateful to anonymous reviewers for many helpful commnets which helped in improving the presentation of this work.


\appendix


\begin{thebibliography}{02}
{\scriptsize
\bibitem{ANLP}
Noga Alon, Amok Bar-Noy, Nathan Linial and David Peleg.
{\em A lower bound for radio broadcast},
Journal of Computer and System Sciences 43(2): 290-298 (1991).

\bibitem{A}
Baruch Awerbuch.
{\em A new distributed depth-first-search algorithm},
Information Processing Letters(1985) 20: 147-150.

\bibitem{BGI}
Reuven Bar-Yehuda, Oded Goldreich and Alon Itai.
{\em On the time complexity of broadcast in radio networks: an exponential gap between determinism and randomization},
Journal of Computer and System Sciences(1992) 45: 104-126.

\bibitem{BGIE}
{\em Errata regarding "On the Time-Complexity of Broadcast in Radio
Networks: An Exponential Gap Between Determinism and Randomization"},
Dec. 2002, available from
$http://www.wisdom.weizmann.ac.il/~oded/p_bgi.html$

\bibitem{BV04}
Carlos Brito and Shailesh Vaya.
{\em Improved lower bound for deterministic broadcasting in radio networks},
In Theoretical Computer Science 412(29): 3568-3578 (2011)

\bibitem{BP}
B. Bruschi and M. Del Pinto.
{\em Lower bounds for the broadcast problem in mobile radio networks},
Distributed Computing 10 (1997), 129-135.

\bibitem{CF}
I. Chlamtac and A. Farago.
{\em Making transmission schedule immune to topology changes in multi-hop packet radio networks},
IEEE/ACM Trans. on Networking 2 (1994), 23-29.

\bibitem{CGGPR02}
B. S. Chlebus, L. Gasieniec, A. Gibbons, A. Pelc and W. Rytter.
{\em Deterministic broadcasting in unknown radio networks},
In Distributed Computing 15 (2002), 27-28.

\bibitem{CGL02}
M. Christersson, L. Gasieniec, and A. Lingas.
{\em Gossiping with bounded size messages in ad-hoc radio networks},
In Proceedings of Twenty Nineth International Colloquium on Automata, Languages and Programming, ICALP 2002, LNCS 2380, 377-389.

\bibitem{CGOR00}
Bogdan S. Chlebus, Leszek Gasieniec, Alan Gibbons, Anna Ostlin and John M. Robson.
{\em Deterministic radio broadcasting.}
In Proceedings 27th International Colloquium on Automata, Languages and Programming, ICALP 2000, LNCS 1853, 717-728.

\bibitem{CGR02}
Marek Chrobak, Leszek Gasieniec and Wojciech Rytter.
{\em Fast broadcasting and gossiping in radio networks},
In Proceedings 41st Annual IEEE Symposium on Foundations of Computer Science, FOCS 2000, 575-581.

\bibitem{CK},
B.S. Chlebus and Dariusz Kowalski.
{\em A better wake-up in radio networks},
In Proceedings of ACM Annual Symposium on Principles of Distributed Computing, PODC 2004.

\bibitem{CMS}
Andrea E.F. Clementi, Angelo Monti and Riccardo Silvestri.
{\em Selective families, superimposed codes, and broadcasting on unknown radio networks},
In Proceedings 12th Annual ACM-SIAM Symposium on Discrete Algorithms, SODA 2001, 709-718.

\bibitem{CR}
Artur Czumaj and W. Rytter.
{\em Broadcasting Algorithms in Radio Networks with Unknown Topology},
In Proceedings of the 44th Annual IEEE Symposium on Foundations of Computer Science, FOCS 2003, Cambridge, MA.

\bibitem{CRM}
B.S. Chlebus and M. Rokicki.
{\em Asynchronous broadcasting in radio networks},
To appear in International Colloquium in Structural Information and Communication Complexity, SIROCCO 2004, LNCS 3104, 57-68.

\bibitem{CRUZ}
R. Cruz and B. Hajek,
{\em A new upper bound to the throughput of a multi-access broadcast channel},
IEEE Transactions on Information Theory IT-28, 3(May 1982), 402-405.


\bibitem{CW}
I. Chlamtac and O. Weinstein.
{\em The wave expansion approach to broadcasting in multihop radio networks},
IEEE Transactions on Communications 39 (1991), 426-433.

\bibitem{EFF}
P. Erdos, P. Frankl, Z. Furedi.
{\em Families of finite sets in which no set is covered by the union of $r$ others},
In Israel Journal of Mathematics, 51 (1985), 79-89.

\bibitem{EK}
M. Elkin and G. Kortsartz.
{\em Improved broadcast schedule for radio networks},
In Sixteenth Annual ACM-SIAM Symposium on Discrete Algorithms, SODA 2005.

\bibitem{FP08}
E. G. Fusco and A. Pelc.
{\em Acknowledged broadcasting in ad-hoc radio networks},
In Information Processing Letters, 109 (2008), 136-141.

\bibitem{G09}
L. Gasienec,
{\em On effecient gossiping in radio networks},
In Proceedings of Sixteenth International Colloquium on Structural Information and Communication Complexity, SIROCCO 2009, LNCS 5869, 2-14.
June 7-11, Şirince, Turkey.

\bibitem{GL02}
L. Gasieniec and A. Lingas,
{\em On adaptive deterministic gossiping in ad hoc radio networks},
In Information Processing Letters 83 (2002), 89-94.

\bibitem{M10}
Gianluca De Marco.
{\em Distributed Broadcast in Unknown Radio Networks},
In SIAM Journal of Computing 39(6) (2010), 2162-2175.

\bibitem{MP01}
Gianluca De Marco and Andrzej Pelc.
{\em Faster broadcasting in unknown radio networks},
In Information Processing Letters 79 (2001), 53-56.

\bibitem{GM}
I. Gabour and Y. Mansour.
{\em Broadcast in radio networks},
In Proceedings of 6th Annual ACM-SIAM Symposium on Discrete Algorithms, SODA 1996, 577-585.

\bibitem{GP}
L. Gasieniec and I. Potapov,
{\em Gossiping with unit size messages in known Radio Networks},
In Proceedings of Second International Conference on Theoretical Computer Science (TCS 2002), Volume 223 IFIP Conference Proceedings, 193-205.

\bibitem{GPP}
L. Gasieniec, A. Pelc and D. Peleg.
{\em The wakeup problem in synchronous broadcast systems},
SIAM Journal on Discrete Mathematics 14, 2001, 207-222.

\bibitem{GPP02}
L. Gasieniec, A. Pagourtzis, and I. Potapov.
{\em Deterministic communication in radio networks with large labels},
In Proceedings of Tenth Eurpoean Symposium on Algorithms, ESA 2002, LNCS 2461, 512-524.

\bibitem{GRX04}
L. Gasieniec, T. Radzik, and Q. Xin.
{\em Faster Deterministic Gossiping in Directed Ad Hoc Radio Networks},
In Scandian Workshop on Algorithmic Theory, SWAT 2004, 397-407, 2004.


\bibitem{H95}
F. K. Hwang.
{\em The time complexity of deterministic broadcast in radio networks},
In Discrete Applied Mathematics 60(1995), 219-222.


\bibitem{JS}
T. Jurdzinski and G. Stachowiak,
{\em Probabilistic algorithms for the wake up problem in single-hop radio networks},
In Proceedings of 13th International Symposium on Algorithms and Computing, ISAAC 2002, LNCS 2518, 525-549.


\bibitem{KM}
Eyal Kushilevitz and Yishay Mansour,
{\em An $\Omega(D\log(N/D))$ lower bound for broadcast in radio networks},
In SIAM Journal of Computing 27 (1998), 702-712

\bibitem{KP04}
Dariusz R. Kowalski and Andrzej Pelc,
{\em Time of deterministic broadcasting in radio networks with local knowledge},
In SIAM Journal of Computing 33 vol 4,(2004): 870-891.

\bibitem{KP2}
Dariusz R. Kowalski and Andrzej Pelc,
{\em Faster Deterministic Broadcasting in Ad Hoc Radio Networks},
In Proceedings 43rd Symposium Theoretical Aspects of Computer Science, STACS 2003.

\bibitem{KP3}
Dariusz R. Kowalski and Andrzej Pelc,
{\em Time complexity of radio broadcasting: adaptiveness vs obliviousness and vs randomization vs determinism},
In Theoretical Computer Science 33 vol 3 (2005): 355-371.

\bibitem{KP4}
Dariusz R. Kowalski and Andrzej Pelc,
{\em Centralized deterministic broadcasting in undirected multihop radio networks},
In Proceedings of 7th International Workshop on Approximation algorithms for Combinatorial Optimization Problems, LNCS 3122(2004): 171-182.

\bibitem{KP5}
Dariusz R. Kowalski and Andrzej Pelc,
{\em Optimal Deterministic Broadcasting in Known Topology Radio Networks},
In Distributed Computing 19 vol 3 (2007): 185-195.

\bibitem{KP6}
Dariusz R. Kowalski and Andrzej Pelc,
{\em Broadcasting in undirected ad hoc radio networks},
In Distributed Computing 18 vol 1(2005): 43-57.

\bibitem{KS}
W.H.Kauz and R.R.C. Singleton,
{\em Nonrandom binary superimposed codes},
In IEEE Transactions on Information Theory 10 (1964): 363-377.

\bibitem{LP}
D. Liu and M. Prabhakaran,
{\em On Randomized Broadcasting and Gossiping in Radio Networks},
In Proceedings of Eigth Annual International Conference on Computing and Combinatorics, COCOON, LNCS 2386(2002), 340-349.

\bibitem{MX}
F. Manne and Q. Xin,
{\em Optimal Gossiping with Unit Size Messages in Known Topology Radio Networks},
In Combinatorial and Algorithmic Aspects of Networking, CAAN, LNCS 2006: 125-135.

\bibitem{OCW03}
T. Okuwa, W. Chen, K. Wada,
{\em An optimal algorithm for acknowledged broadcasting in ad hoc radio networks},
In Proceedings of Second International Symposium on Parallel and Distributed Computing (2003), 178-184.

\bibitem{P}
D. Peleg,
{\em Deterministic radio broadcast with no topological knowledge},
Manuscript, 2000.

\bibitem{UCW07}
Jiro Uchida, Wei Chen, Koichi Wada,
{\em Acknowledged broadcasting and gossiping in ad hoc radio networks},
In Theoretical Computer Science, 377 (2007): 43-54.

\bibitem{V11}
S. Vaya,
{\em Faster Gossiping in Bidirectional Radio Networks with Large Labels},
In http://arxiv.org/, 2011.

\bibitem{Xu}
Xu, Y.,
{\em An $O(n^{3/2})$ deterministic algorithm for radio networks},
In Algorithmica 36 vol 1(2003): 93-96.
}
\end{thebibliography}
\end{document}